\documentclass[conference]{IEEEtran}

\usepackage[T1]{fontenc}
\usepackage[utf8]{inputenc}
\usepackage{graphicx}
\usepackage{booktabs}
\usepackage{amsmath}
\usepackage{siunitx}
\usepackage{url}
\usepackage{multirow}
\usepackage{xcolor}
\usepackage{makecell}
\usepackage[hidelinks]{hyperref}
\usepackage{array}

\usepackage{dblfloatfix}
\usepackage{caption}

\usepackage{tikz}
\usetikzlibrary{arrows.meta, positioning, fit, calc}

\title{Energy-Aware LoRaWAN Design for Long-Lived Agricultural Sensing: Insights from a Multi-Year Deployment and Controlled Platform Comparison}

\author{
\IEEEauthorblockN{Carl Dickinson, Shishir Nagaraja, Chuadhry Mujeeb Ahmed}
\IEEEauthorblockA{
School of Computing, Newcastle University\\
National Edge AI Hub\\
Newcastle upon Tyne, UK\\
Email: \{carl.dickinson, shishir.nagaraja, mujeeb.ahmed\}@ncl.ac.uk
}
}

\begin{document}

\maketitle

\begin{abstract}
Agricultural IoT nodes are often far from mains power and have limited access for maintenance. Therefore, IoT deployments benefit from low-power, long-range communication such as LoRaWAN. However, communication alone does not guarantee multi-year operation. Previous work has measured communication energy, modelled system lifetime, and documented long-lived deployments. This paper connects long-term deployment evidence with controlled platform measurements. The AgriTrust deployment has operated for 2 years and 8 months using MKR WAN1310-based Squirrel Box nodes. Building on this evidence, we compared the MKR WAN1310 and STM32WL LoRaWAN platforms using time-resolved current traces during confirmed-uplink cycles. For this outdoor comparison, we tested nine combinations of payload size and data rate with three repetitions each. Both platforms achieved 3/3 ACKed packets in every condition. The STM32WL had lower cycle energy than the WAN1310 in all tested conditions. Cycle energy ranged from $0.0350$~J to $0.3550$~J for STM32WL and from $0.0698$~J to $0.5662$~J for WAN1310. These communication events are brief, while sleep and leakage currents persist between cycles. The STM32WL sleep current was recorded at $59.6$~nA, while the WAN1310 sleep current was $104$~$\mu$A, approximately $1745\times$ that of the STM32WL. This difference shows that sleep current is a first-order determinant of long-term viability. However, the whole-system energy budget also depends on sensing loads, duty cycle, communication energy, and harvested-energy support. The study should be treated as a controlled characterisation, not a universal comparison across all sites, firmware stacks, or LoRaWAN modes.
\end{abstract}

\section{Introduction}
Agricultural IoT nodes often operate far from mains power, with intermittent connectivity and limited maintenance access. LoRaWAN is attractive because it provides long range and low average power, but multi-year operation is not explained by communication technology alone. Endurance also depends on sleep architecture, sensing schedule, power-tree overheads, and energy harvesting \cite{casals2017,bouguera2018,maudet2021,raghunathan2005,kansal2007}.

This paper addresses a gap between two common forms of evidence. Existing work provides analytical and measurement-based models of LoRaWAN communication energy \cite{casals2017,bouguera2018,maudet2021,thoen2019}, while agricultural deployments often demonstrate utility without exposing the board-level energy mechanisms that support long lifetime \cite{codeluppi2020,singh2020greenhouse,novak2025}. Less common is a combined treatment of real deployment evidence, controlled platform comparison, and system-level lifetime interpretation.

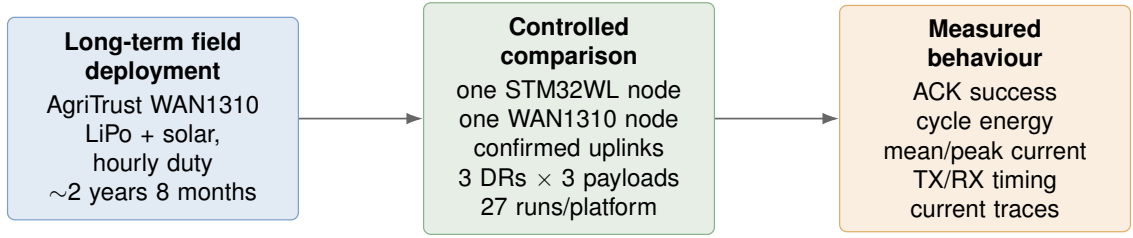
\begin{figure*}[t]
\centering
\resizebox{0.82\textwidth}{!}{%
\begin{tikzpicture}[
    font=\sffamily,
    node distance=12mm and 18mm,
    >=Latex,
    mainbox/.style={
        rounded corners=3pt,
        draw=black!45,
        line width=0.5pt,
        align=center,
        inner sep=6pt,
        minimum height=18mm,
        text width=38mm
    },
    arrow/.style={
        -{Latex[length=2.4mm,width=1.7mm]},
        line width=0.7pt,
        draw=black!60
    },
    grouplabel/.style={
        font=\sffamily\scriptsize,
        align=center,
        fill=white,
        inner xsep=1.5pt,
        inner ysep=0.8pt
    }
]

\definecolor{deployblue}{RGB}{64,118,180}
\definecolor{measuregreen}{RGB}{85,145,90}
\definecolor{energyorange}{RGB}{215,145,50}

\node[mainbox, fill=deployblue!15, draw=deployblue!70] (deployment) {
    \textbf{Long-term field\\deployment}\\[2pt]
    AgriTrust WAN1310\\
    LiPo + solar, hourly duty\\
    $\sim$2 years 8 months
};

\node[mainbox, fill=measuregreen!15, draw=measuregreen!70, right=of deployment] (experiment) {
    \textbf{Controlled\\comparison}\\[2pt]
    one STM32WL node\\
    one WAN1310 node\\
    confirmed uplinks\\
    3 DRs $\times$ 3 payloads\\
    27 runs/platform
};

\node[mainbox, fill=energyorange!15, draw=energyorange!75, right=of experiment] (metrics) {
    \textbf{Measured\\behaviour}\\[2pt]
    ACK success\\
    cycle energy\\
    mean/peak current\\
    TX/RX timing\\
    current traces
};

\draw[arrow] (deployment.east) -- (experiment.west);
\draw[arrow] (experiment.east) -- (metrics.west);

\end{tikzpicture}%
}

\vspace{0.5ex}

\caption{Deployment-informed LoRaWAN energy evaluation framework linking the long-running AgriTrust deployment with the controlled STM32WL--WAN1310 measurement workflow.}
\label{fig:paper_overview}
\end{figure*}

We make four contributions. First, we use the AgriTrust deployment as a real-world anchor for long-lived agricultural sensing \cite{dickinson2023agritrust}. Secondly, using one instrumented STM32WL node and one instrumented WAN1310 node, controlled confirmed-uplink experiments show that both platforms complete the outdoor comparison matrix: DR0/DR3/DR5, 15/30/45~B payloads, three repetitions per condition, and 27/27 ACKed messages per platform. STM32WL consistently achieves lower cycle energy, shorter radio-active time, and lower mean current than WAN1310 in this setup. Thirdly, we report a measured sleep-current disparity of approximately 1745$\times$. Fourthly, we show that communication energy alone cannot explain multi-year lifetime: sleep current and system design dominate the long-term outcome. Figure~\ref{fig:paper_overview} summarises how the deployment and measurements connect.

\section{Background and Related Work}
LoRaWAN Class~A energy is tightly coupled to MAC timing. Each uplink is followed by receive windows, so even when application downlinks are absent there remains a deterministic post-uplink listening cost \cite{lorawan104,lorawanrp002}. Confirmed traffic further couples energy to ACK reception and possible retransmission behaviour, while data-rate (DR) selection changes airtime through the regional spreading-factor and bandwidth mapping \cite{lorawan104,lorawanrp002,semtech2013}. Measurement-based work shows that realised cost is not set by airtime alone: receive-window overheads, retries, firmware behaviour, and board-level implementation can all affect total message energy \cite{casals2017,bouguera2018,maudet2021,ould2021boards}.

At system level, long lifetime is usually estimated from cycle energy, daily demand, and available battery or harvested energy \cite{bouguera2018,thoen2019,singh2020energy}. However, practical LoRa sensing studies show that sensing load, reporting interval, sleep-mode current, standby leakage, and application scheduling can dominate the budget once reporting becomes sparse \cite{vaananen2022compression,singh2020energy,novak2025}. This matters especially in agriculture and environmental monitoring, where remote placement, link variability, and maintenance cost are normal rather than exceptional \cite{codeluppi2020,singh2020greenhouse,liopa2024,perret2025}.

Solar assistance shifts the design objective from finite battery lifetime to maintaining a positive long-term energy balance under variable harvest \cite{raghunathan2005,kansal2007,peng2014,larosa2024}. Recent LoRa-based environmental-sensing work further shows that fixed duty-cycle assumptions may fail under changing harvest conditions, motivating adaptive scheduling and system-level energy management \cite{ma2024solar}. Taken together, prior work provides strong models, measurements, and deployments, but fewer studies combine long-horizon agricultural evidence, controlled cross-platform LoRaWAN measurements, and system-level interpretation of sleep leakage, sensing load, duty cycle, and harvested-energy support.

\section{Deployment Context and Research Questions}
AgriTrust provides the deployment context for this study \cite{dickinson2023agritrust}. The node is based on the Arduino MKR WAN 1310, powered by a \SI{3.7}{\volt} LiPo battery with solar charging and an external timer-assisted wake/sleep strategy. Its duty cycle is approximately hourly, and project records indicate a field duration of roughly 2 years and 8 months.

The deployment shows that long-lived agricultural sensing is a system-design problem rather than simply a radio-efficiency problem. Deep sleep between events, controlled wake duration, sparse reporting, and harvested-energy support make long operation possible. This motivates the controlled comparison used in this paper: if a platform is field-capable, what is its board-level energy behaviour under matched confirmed-uplink experiments?

We therefore ask how reliably the two platforms execute confirmed uplink cycles, how energy, current, and timing vary with data rate and payload, whether cleaner board-level behaviour makes one platform more suitable for low-power experimentation, and how communication measurements should be interpreted alongside deployment lifetime, solar support, and sleep current.

\section{Experimental Method and Datasets}
The experiments use a host-driven LoRaWAN test workflow: a PC-side script configures each run, starts current capture, coordinates the node-side action, and stores the run summary, validation output, current trace, host log, and device event log in a per-run directory. This preserves the electrical trace and event timing needed to separate wake, TX, RX, and acknowledgement behaviour.

We focus on confirmed uplinks because they expose transmission, LoRaWAN Class~A receive windows, and acknowledgement (ACK) success. In a confirmed uplink, the device transmits a frame requesting confirmation, opens receive windows after TX, and the exchange is successful only if the network ACK is observed. Thus \texttt{missing\_ack\_warning} means the run was measured and parsed, but the expected ACK was not observed. The reported results therefore reflect reliable-delivery cost under this workflow; unconfirmed uplink was not evaluated. Here, data rate (DR) denotes the region-specific PHY setting, with lower DRs implying longer airtime.

One run corresponds to one transmission attempt and at most one packet. A condition is one platform/DR/payload combination with three outdoor repetitions; condition-level metrics are the aggregated values over those repetitions, such as mean cycle energy/current/timing and ACK totals such as 3/3.

The two main experimental platforms are the ST B-WL5M-SUBG1 STM32WL board and WAN1310, both operated from a \SI{3.7}{\volt} supply during the reported runs. The workflow records time-resolved current traces together with host and device event logs, allowing each run to be analysed in terms of cycle energy, mean current, peak current, and phase timing. The main outdoor comparison uses an approximately \SI{100}{\meter} outdoor distance path and a matched matrix of three data rates (DR0, DR3, DR5), three payloads (15~B, 30~B, 45~B), and three repetitions per condition, giving 27 runs per outdoor dataset. The analysed STM32WL outdoor dataset records \texttt{tx\_periodicity\_ms=5000}, while the analysed WAN1310 outdoor dataset records \texttt{tx\_periodicity\_ms=150000}. This preserves the same communication matrix while allowing platform-appropriate inter-run spacing.

The longer WAN1310 interval follows Arduino's documented MKRWAN modem limit of one message every two minutes \cite{arduino_mkrwan_lorasendreceive_tutorial}. Using \SI{150}{\second} keeps WAN1310 above that firmware-imposed limit while preserving the same payload and data-rate matrix as STM32WL. The comparison is therefore a controlled per-cycle cost comparison, not a throughput comparison under identical inter-run spacing. Retransmission behaviour is not fully observable, so the analysis is based on observed cycle completion and ACK reception.

Table~\ref{tab:sysconfig} summarises the configuration details that can be confirmed directly from the code, metadata, and run summaries used in this paper.

\begin{table}[htbp]
\caption{System configuration.}
\label{tab:sysconfig}
\centering
\renewcommand{\arraystretch}{1.15}
\begin{tabular}{p{2.45cm}p{5.1cm}}
\toprule
Item & Value confirmed from artefacts \\
\midrule
Region & EU868 \\[4pt]
LoRaWAN class & Class~A for STM32WL firmware; WAN1310 class not explicitly logged, default stack behaviour \\[4pt]
Traffic type & Confirmed uplink \\[4pt]
ADR state & Disabled (\texttt{adr\_enable=0}) in the analysed runs \\[4pt]
Data rates used & DR0, DR3, DR5 \\[4pt]
Payload sizes & 15~B, 30~B, 45~B \\[4pt]
Repetitions & 3 per condition where available \\[4pt]
Cadence in analysed runs & STM32WL outdoor dataset: \texttt{tx\_periodicity\_ms=5000}; WAN1310 outdoor dataset: \texttt{tx\_periodicity\_ms=150000} \\[4pt]
TX power & STM32WL run summaries log 13~dBm; WAN1310 TX power was not surfaced by the saved experiment artefacts or configuration snapshots \\[4pt]
Bandwidth & 125~kHz in logged radio-configuration events/run summaries \\[4pt]
Coding rate & STM32WL run summaries log CR4/5; WAN1310 coding rate follows stack/modem behaviour and was not surfaced by the saved experiment artefacts \\[4pt]
Frequency & STM32WL run summaries log EU868 uplink frequencies; WAN1310 frequency follows stack/modem channel selection and was not surfaced by the saved experiment artefacts \\[4pt]
Gateway model/configuration & WisGate Edge Pro, model RAK7289CV2, using built-in server configuration \\[4pt]
Network server & Built-in gateway server; Network ID 1; server-side ADR enabled with ADR margin 5~dB; minimum allowed uplink DR0 (SF12/BW125), maximum allowed uplink DR7 (FSK 50~kbps); downlink TX power 20~dBm \\[4pt]
RX1/RX2 parameters & RX1 delay 1~s and RX1 data-rate offset 0; RX2 frequency 869.525~MHz at DR0 (SF12/BW125) \\[4pt]
Retrans. settings & Confirmed uplink used; run summaries report \texttt{retry\_count=0} in analysed examples; frame-counter validation disabled in the gateway server configuration \\[4pt]
Current measurement device & Current Ranger \\
\bottomrule
\end{tabular}
\renewcommand{\arraystretch}{1}
\end{table}

Table~\ref{tab:datasets} summarises the experimental datasets used in the paper.

\begin{table}[htbp]
\caption{Datasets used in the paper.}
\label{tab:datasets}
\centering
\small
\setlength{\tabcolsep}{4pt}
\begin{tabular}{
>{\centering\arraybackslash}m{1.25cm}
c
>{\centering\arraybackslash}m{1.0cm}
c
c
>{\centering\arraybackslash}m{1.35cm}
}
\toprule
Dataset & Board & Setting & Runs & Msgs. & Use \\
\midrule
\makecell{DR\\sweep} & STM32WL & \makecell{pre-\\cursor} & 8 & 8 & \makecell{Prelim.\\trend} \\
\makecell{Outdoor} & STM32WL & outdoor & 27 & 27 & \makecell{Main\\comparison} \\[12pt]
\makecell{Outdoor} & WAN1310 & outdoor & 27 & 27 & \makecell{Main\\comparison} \\
\bottomrule
\end{tabular}
\end{table}

The STM32WL DR sweep serves as a preliminary characterisation for the outdoor study. It comprises eight valid runs across DR0--DR5 at 15~B and demonstrates a clear airtime-driven reduction in energy consumption, decreasing from DR0 (\SI{0.2561}{\joule}) to DR4 (\SI{0.0399}{\joule}) and DR5 (\SI{0.0325}{\joule}), consistent with expected communication costs. The primary quantitative comparison between platforms in this work is based exclusively on the two outdoor datasets.

\section{Results}

\subsection{Reliability}
In the analysed outdoor comparison, both platforms produce a complete matrix: all 27 STM32WL runs and all 27 WAN1310 runs validate successfully, all nine DR/payload conditions produce packets, and every condition achieves 3/3 ACKed packets. Given the uniform ACK performance across all conditions, the analysis emphasises energy and current measurements as the distinguishing metrics.

\begin{figure*}[htbp]
\centering
\includegraphics[width=0.95\textwidth]{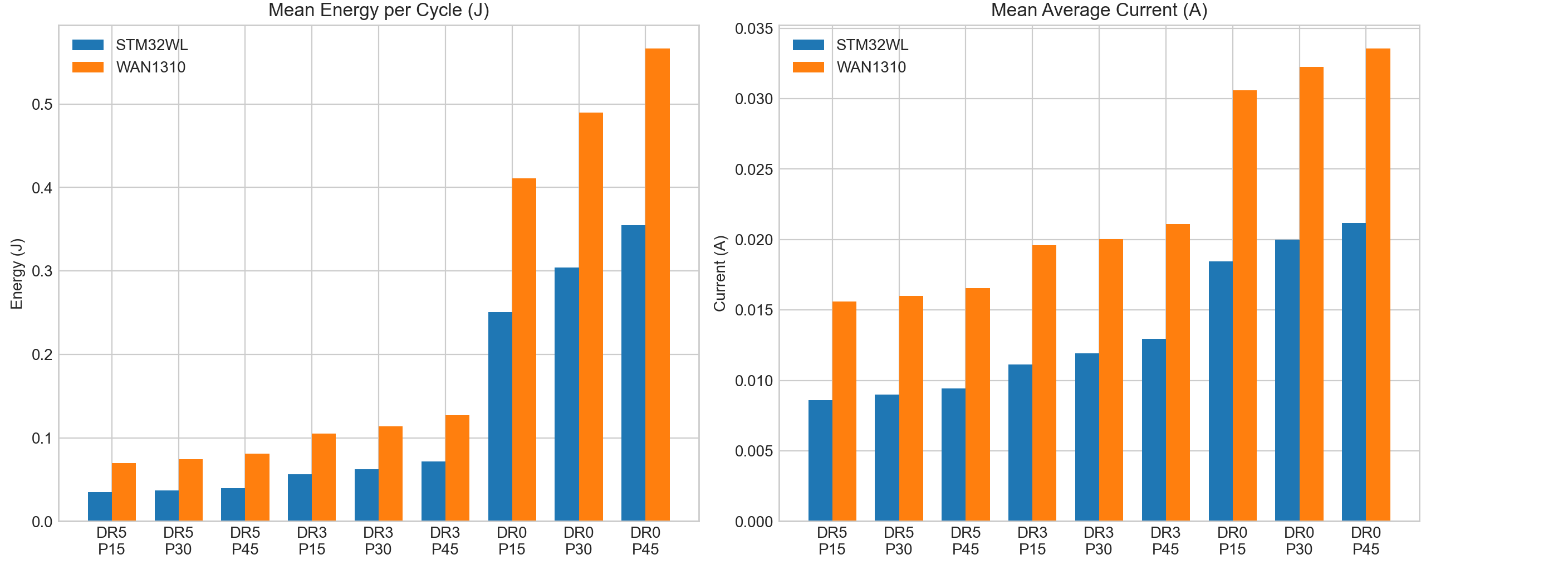}
\caption{Core outdoor distance comparison. All outdoor conditions achieved 3/3 ACKed packets on both platforms. Bars show three-run condition means; maximum coefficient of variation (CV) is 2.4\% for energy and 2.2\% for mean current, so error bars are omitted for readability. STM32WL remains lower in energy and mean current across the matrix.}
\label{fig:core}
\end{figure*}

\subsection{Energy Behaviour}
For the STM32WL outdoor dataset, mean cycle energy is consistent and physically interpretable across the full matrix. At DR5, cycle energy rises gently with payload from \SI{0.0350}{\joule} at 15~B to \SI{0.0397}{\joule} at 45~B. At DR3, the corresponding values are \SI{0.0563}{\joule}, \SI{0.0624}{\joule}, and \SI{0.0714}{\joule}. At DR0, cycle energy rises further to \SI{0.2505}{\joule}, \SI{0.3037}{\joule}, and \SI{0.3550}{\joule}, reflecting the much longer airtime and receive-window occupancy at the lowest data rate.

The WAN1310 outdoor dataset shows the same qualitative airtime-driven trend but at a consistently higher absolute cost. At DR5, mean cycle energy rises from \SI{0.0698}{\joule} at 15~B to \SI{0.0811}{\joule} at 45~B. At DR3, the values are \SI{0.1053}{\joule}, \SI{0.1135}{\joule}, and \SI{0.1274}{\joule}. At DR0, they rise to \SI{0.4106}{\joule}, \SI{0.4894}{\joule}, and \SI{0.5662}{\joule}. Because all outdoor runs are successful confirmed uplinks in the analysed datasets, energy per successful packet is identical to cycle energy for both platforms throughout the matrix.

The key comparison is therefore straightforward. Both platforms exhibit coherent scaling with data rate and payload, but STM32WL remains lower in cycle energy in every outdoor condition. The gap is already about \SI{2.0}{\times} at DR5/15~B (\SI{0.0350}{\joule} versus \SI{0.0698}{\joule}) and remains substantial at DR0/45~B (\SI{0.3550}{\joule} versus \SI{0.5662}{\joule}). The longer \SI{150}{\second} WAN1310 inter-run spacing does not contribute to these values because the reported energy is integrated over the active measurement window of each confirmed-uplink cycle rather than over the idle time between runs.

\subsection{Timing and Current Behaviour}
\begin{figure*}[htbp]
\centering
\includegraphics[width=0.88\textwidth]{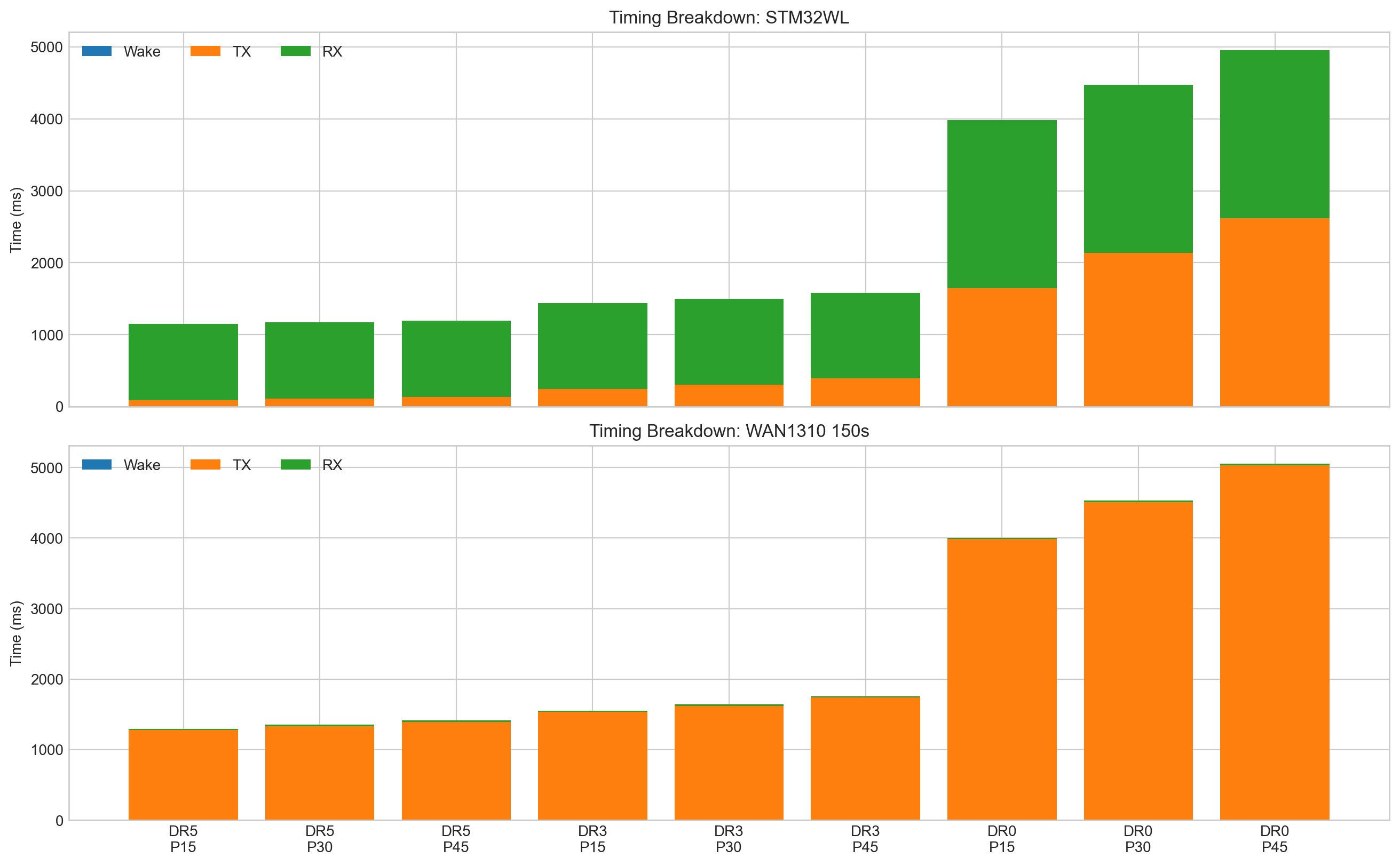}
\caption{Timing breakdown for the outdoor distance comparison. Bars show three-run condition means; maximum coefficient of variation is 1.2\% across plotted timing components, so error bars are omitted for readability. Both platforms show airtime-driven growth as data rate decreases and payload increases. RX timing should be interpreted carefully because STM32WL and WAN1310 expose receive behaviour at different instrumentation layers.}
\label{fig:timing}
\end{figure*}

Figure~\ref{fig:timing} explains the energy trends mechanistically. For STM32WL, wake time is nearly constant at about \SI{7.1}{\milli\second}, while TX and RX dominate the cycle. TX time expands from \SI{83.6}{\milli\second} at DR5/15~B to \SI{2611.1}{\milli\second} at DR0/45~B; RX time rises from about \SI{1058}{\milli\second} at DR5 to about \SI{2337}{\milli\second} at DR0. The total cycle duration therefore grows from roughly \SI{1.21}{\second} at DR5/15~B to about \SI{5.02}{\second} at DR0/45~B.

For WAN1310, the timing picture is also internally consistent across the matrix, but the radio-active portion is longer. TX time grows from \SI{1278.6}{\milli\second} at DR5/15~B to \SI{5033.7}{\milli\second} at DR0/45~B, and total cycle duration grows from \SI{1.35}{\second} to \SI{5.11}{\second}. A notable platform-level difference is that WAN1310 shows a very short logged RX interval of about \SI{20}{\milli\second} across the matrix, whereas STM32WL exposes a much longer receive-window contribution in the saved run summaries. This should not be over-interpreted as a literal statement that the WAN1310 radio physically listens for only \SI{20}{\milli\second}. Rather, the two boards expose RX timing at different instrumentation layers. In the STM32WL firmware, \texttt{EV\_RX\_WAIT\_START} is emitted when the stack enters its receive phase and \texttt{EV\_RX\_DONE} is emitted when that phase closes, so the logged \texttt{rx\_time\_ms} includes the receive-window wait. In the WAN1310 sketch, by contrast, \texttt{EV\_RX\_WAIT\_START} is emitted only after \texttt{modem.endPacket(...)} returns, so the saved \texttt{rx\_time\_ms} reflects only the short sketch-visible completion interval between that return point and the subsequent ACK/RX-done events. The mean average-current bars in Fig.~\ref{fig:core} reinforce the same point: STM32WL average current ranges from \SI{8.59}{\milli\ampere} to \SI{21.16}{\milli\ampere}, while WAN1310 ranges from \SI{15.61}{\milli\ampere} to \SI{33.54}{\milli\ampere}. Peak current is also generally higher for WAN1310, reaching about \SI{55.2}{\milli\ampere} at DR5/15~B compared with \SI{33.3}{\milli\ampere} for STM32WL.

\begin{figure*}[htbp]
\centering
\includegraphics[width=0.88\textwidth]{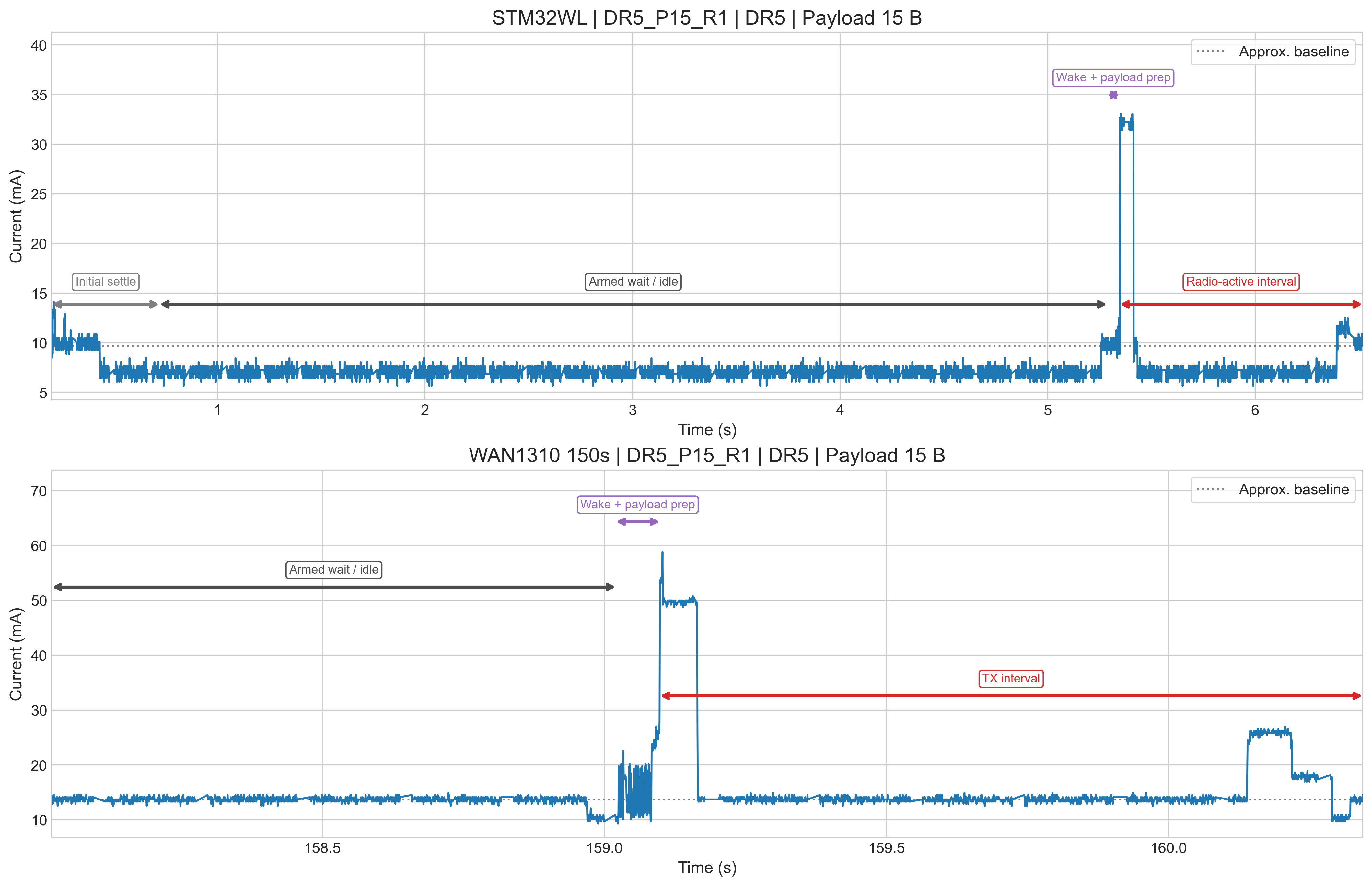}
\caption{Representative single-run current traces from complete successful cycles; standard deviations are therefore not plotted. WAN1310 remains longer and more bursty during radio activity than STM32WL.}
\label{fig:traces}
\end{figure*}

Figure~\ref{fig:traces} provides a trace-level view of the same behaviour. The STM32WL representative trace shows a clearly segmented cycle with a stable idle period followed by a compact radio-active interval. The representative WAN1310 trace, by contrast, contains a longer radio-active interval with multiple separated current bursts. This is useful scientifically because it links the aggregate timing and energy metrics back to observable current behaviour.

\subsection{System vs Platform Energy}
The controlled experiments are intentionally platform-centric rather than full deployment replicas. They measure the electrical cost of the communication cycle under a matched workflow, allowing board behaviour to be compared without implying that the resulting joule values are the total daily energy of an agricultural node. This distinction matters because sensing, storage, wake circuitry, battery characteristics, and solar charging all sit outside the radio-only budget \cite{bouguera2018,thoen2019,novak2025}.

Our measurements nevertheless make one useful system-level point. Even the heaviest successful outdoor condition in the controlled comparison, WAN1310 DR0/45~B at \SI{0.5662}{\joule}, remains a sub-joule event. Communication is therefore episodic, whereas long-term survival depends mainly on what happens between cycles: sleep leakage, sensing frequency, and whether harvested energy can cover the average daily demand.

At system level, a first-order energy budget can be written as the sum of the dominant tasks within one reporting cycle:
\begin{equation}
\begin{aligned}
E_{\text{total}} \approx {} & E_{\text{MCU,on}} + E_{\text{sensors,power}} + E_{\text{sensing}} \\
& + E_{\text{processing}} + E_{\text{storage}} + E_{\text{comm}} + E_{\text{sleep}} .
\end{aligned}
\end{equation}
Here, $E_{\text{comm}}$ corresponds to the cycle energy characterised in this paper, while the remaining terms depend on the sensing stack, processing workload, storage behaviour, and duty cycle.

Local processing is captured by $E_{\text{processing}}$. It may be negligible for thresholding or packet formatting, but can become material for embedded AI. As an indicative STM32WL-class measurement, a quantised WISDM activity-recognition model deployed through X-CUBE-AI consumed approximately \SI{2.59}{\joule} over 400 logged samples, assuming a \SI{3.7}{\volt} supply, or about \SI{6.48}{\milli\joule} per logged sample. Although outside the LoRaWAN comparison dataset, this illustrates why AI workflows, especially camera vision, must be treated as additional active loads.

For agricultural nodes, the sensing side can dominate: the energy required to power sensors, wait for them to settle, acquire measurements, optionally log data to storage, and then switch those peripherals off can exceed the radio cost. In that sense the full node budget is approximately additive at first order, although real implementations also inherit losses from regulators, switching elements, and battery behaviour \cite{bouguera2018,thoen2019}.

Table~\ref{tab:consumption} is included to illustrate how sensing and storage load can shift the overall energy balance away from radio energy alone. It gives a representative AgriTrust peripheral current budget that is used later for a worked example. Summing these devices gives an active peripheral load of \SI{97.9}{\milli\ampere}. In the worked example below, this load is held constant across platforms so that the effect of the different sleep-current cases can be compared directly.

\begin{table}[!t]
\vspace{-1.0ex}
\caption{Illustrative AgriTrust peripheral current budget based on typical operating currents under an all-active load assumption, not a full measured time-profile.}
\label{tab:consumption}
\centering
\small
\setlength{\tabcolsep}{4pt}
\begin{tabular}{lc}
\toprule
Device & Current \\
\midrule
Soil temperature sensor & 0.7 mA \\
Soil moisture sensors & 6.6 mA \\
pH sensor & 11 mA \\
NPK sensor & 22 mA \\
Ambient light sensor & 7 mA \\
Ambient temperature \& humidity sensor & 0.6 mA \\
SD card module & 50 mA \\
\midrule
Total active peripheral load & 97.9 mA \\
\bottomrule
\end{tabular}
\vspace{-2.0ex}
\end{table}

This framing motivates practical power-saving strategies: switch sensor and storage rails only for the measured time needed to complete each task, return the MCU to its lowest viable sleep state after transmission, and use external timer-based power gating when board-level sleep current remains too high.

\subsection{Sleep Current Impact}
Sleep current is a first-order determinant of long-term behaviour in sparse-duty-cycle operation. The measured contrast is \SI{59.6}{\nano\ampere} versus \SI{104}{\micro\ampere}, or about 1745$\times$. The STM32WL value is plausible against STM32WLE5 datasheet figures of \SI{31}{\nano\ampere} shutdown, \SI{360}{\nano\ampere} standby with RTC, and \SI{1.07}{\micro\ampere} Stop2 with RTC at \SI{3}{\volt} \cite{stm32wle5}. The WAN1310 value matches Arduino's board-level statement that MKR WAN 1310 can reach \SI{104}{\micro\ampere} when properly configured \cite{arduino_mkrwan1310_store}, but remains configuration-dependent \cite{arduino_mkrwan_sleep_current_issue79}. These are measured platform-level sleep figures, not a fully harmonised board-for-board benchmark. Table~\ref{tab:sleep} converts them into idealised battery-only terms for a \SI{3.7}{\volt}, \SI{3700}{\milli\ampere\hour} battery.

The daily sleep-energy term is computed as shown in Table \ref{tab:sleep}.
\begin{table}[htbp]
\caption{Idealised impact of sleep current for a \SI{3.7}{\volt}, \SI{3700}{\milli\ampere\hour} battery.}
\label{tab:sleep}
\centering
\small
\begin{tabular}{lccc}
\toprule
Platform & Sleep current & \makecell{Sleep\\energy/day\\(Wh)} & \makecell{Sleep-only\\lifetime\\(days)} \\
\midrule
STM32WL & \SI{59.6}{\nano\ampere} & 5.29 $\mu$Wh & $2.59\times10^{6}$ \\
WAN1310 & \SI{104}{\micro\ampere} & 9.24 mWh & 1482 \\
\bottomrule
\end{tabular}
\end{table}

These lifetimes are comparative, not deployment predictions: they ignore battery self-discharge, temperature, sensing load, and communication. At 24 transmissions/day, the representative STM32WL DR5/15~B communication cost is only 0.233~mWh/day, whereas a constant \SI{104}{\micro\ampere} sleep current consumes about 9.24~mWh/day. External power-gating can reduce the effective WAN1310 sleep term substantially; AgriTrust-style timer operation can reach about \SIrange{5}{6}{\micro\ampere}, while nano-power timer devices such as the TPL5110 are around \SI{39}{\nano\ampere} \cite{attiny84_datasheet,tpl5110_datasheet}.

\subsection{Lifetime and Solar Interpretation}
For a node executing $N_{\text{cycles}}$ communication cycles per day with energy $E_{\text{cycle}}$, the communication-only daily energy demand is
\begin{equation}
E_{\text{daily}} = E_{\text{cycle}} \times N_{\text{cycles}} .
\end{equation}
Expressing battery capacity in watt-hours, an ideal battery-only lifetime estimate is
\begin{equation}
\text{Lifetime}_{\text{days}} = \frac{\text{Battery}_{\text{Wh}}}{E_{\text{daily}}}.
\end{equation}

\begin{figure}[htbp]
\centering
\includegraphics[width=\columnwidth]{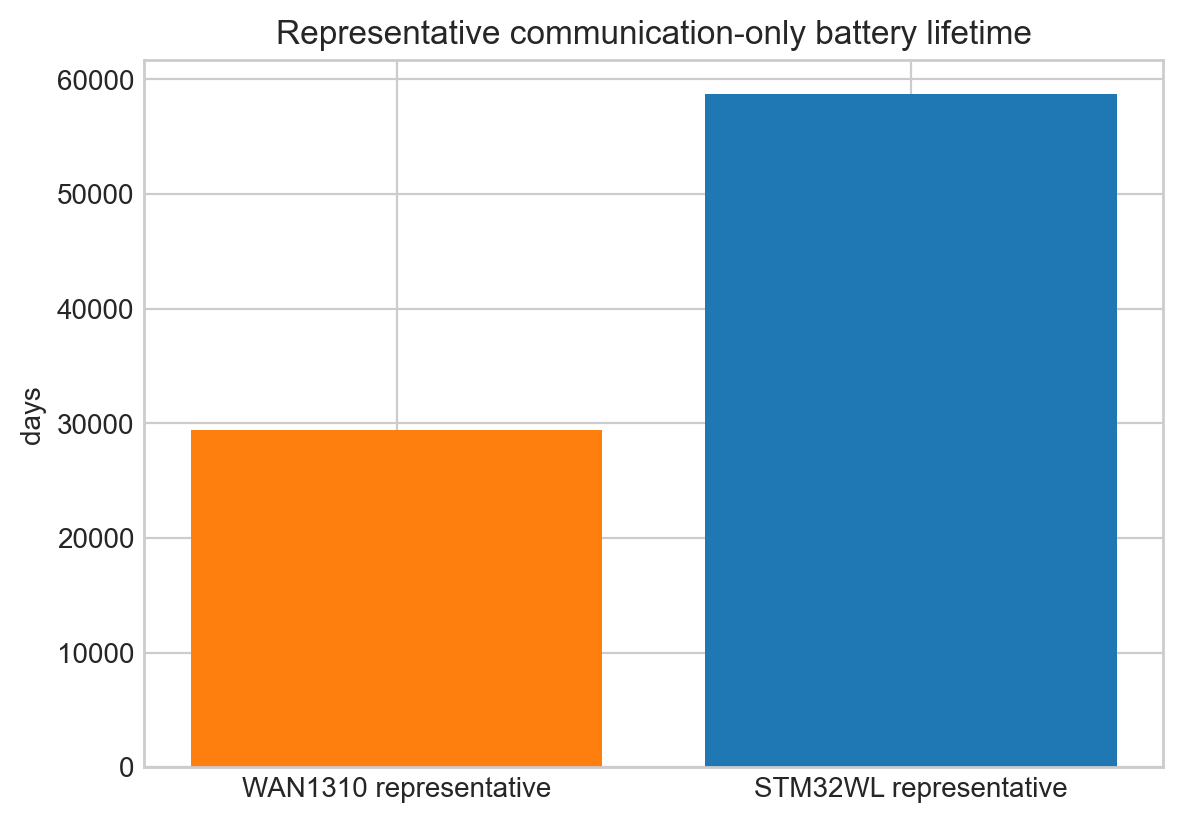}
\caption{Communication-only battery-lifetime comparison for representative outdoor STM32WL and WAN1310 conditions. The figure is illustrative rather than predictive.}
\label{fig:budget}
\end{figure}

We use a \SI{3.7}{\volt}, \SI{3700}{\milli\ampere\hour} battery, 24 cycles/day, and an illustrative solar recharge of 0.6~Wh/day. For representative STM32WL DR5/15~B, $E_{\text{cycle}}=\SI{0.0350}{\joule}$ gives $E_{\text{daily}}\approx 0.000233$~Wh/day. This communication-only result is not a deployment prediction; it shows that radio energy is small relative to the battery and solar budget, so long-term behaviour is governed by sleep leakage, sensing strategy, duty cycle, and harvested-energy support. Figure~\ref{fig:budget} summarises this interpretation.

To extend the communication results to a fuller agricultural-node budget, Table~\ref{tab:agritrust_example} applies the \SI{97.9}{\milli\ampere} shared active load from Table~\ref{tab:consumption}, \SI{10}{\second} active time, \SI{3590}{\second} sleep time, a \SI{3.7}{\volt} \SI{3700}{\milli\ampere\hour} battery (\SI{13.69}{\watt\hour} nominal, 80\% usable), and \SI{1.4}{\watt\hour\per\day} usable solar input from a \SI{0.5}{\watt} panel. It is illustrative, not a re-decomposition of the measured traces. The average current is

\begin{equation}
I_{\text{avg}}=\frac{I_{\text{active}}t_{\text{active}}+I_{\text{sleep}}t_{\text{sleep}}}{t_{\text{cycle}}}.
\end{equation}
and the corresponding full-duty-cycle daily energy is
\begin{equation}
E_{\text{day,avg}} = V_{\text{bat}} \times I_{\text{avg}} \times 24.
\end{equation}
The same active load is applied to each case so the sleep-current effect can be read directly. \emph{Nominal days} uses the full \SI{13.69}{\watt\hour} battery energy, \emph{Usable days} uses the derated 80\% usable battery energy, and positive solar surplus means daily harvest exceeds daily consumption.

\begin{table}[htbp]
\caption{Representative AgriTrust-style lifetime example with shared active load.}
\label{tab:agritrust_example}
\centering
\small
\setlength{\tabcolsep}{2pt}
\begin{tabular}{@{}p{1.2cm}cccccc@{}}
\toprule
\centering Case & \makecell{Sleep\\current} & \makecell{$I_{\text{avg}}$\\(mA)} & \makecell{Daily\\energy\\(Wh/day)} & \makecell{Nominal\\days} & \makecell{Usable\\days} & \makecell{Solar\\surplus\\(Wh/day)} \\
\midrule
\makecell[c]{WAN\\1310} & \SI{104}{\micro\ampere} & 0.376 & 0.0334 & 410 & 328 & +1.3666 \\[12pt]
\makecell[c]{WAN\\1310\\+ timer} & \SI{6}{\micro\ampere} & 0.278 & 0.0247 & 555 & 444 & +1.3753 \\[15pt]
\makecell[c]{STM32\\WL} & \SI{59.6}{\nano\ampere} & 0.272 & 0.0242 & 567 & 453 & +1.3758 \\
\bottomrule
\end{tabular}
\end{table}

This example shows that once sensing and storage loads are included, the system moves from extremely large communication-only lifetimes to a more plausible battery-only range of roughly 1 to 1.5 years. Under the stated solar assumptions, all three cases are energy-positive daily, explaining why AgriTrust-style operation depends on solar balance, battery ageing, environmental conditions, and hardware reliability rather than nominal battery capacity alone.

\section{Discussion}
Two conclusions emerge clearly from the analysed datasets. First, the AgriTrust deployment demonstrates practical viability: a LoRaWAN agricultural node can remain useful over multiple years, but only when the entire system is engineered for low average power. Secondly, the controlled comparison demonstrates clear platform differences even when both boards complete the same outdoor confirmed-uplink matrix successfully: STM32WL is consistently lower in energy, mean current, peak current, and TX duration than WAN1310 across the measured payload and data-rate combinations. The deployment shows that long life is achievable; the controlled comparison shows which platform properties make that outcome easier to engineer and analyse.

This has a broader methodological implication. Comparing platforms by joules alone is most informative when both sides are executing comparable successful transactions; the present outdoor matrix provides exactly that condition. The remaining differences can therefore be attributed more directly to board- and stack-level behaviour rather than to missing acknowledgements or truncated cycles. This is also consistent with prior work suggesting that system energy can be strongly influenced by standby, sensing, and platform overheads rather than by radio transmission alone \cite{bouguera2018,thoen2019,singh2020energy,novak2025}. The deployment and controlled experiments therefore complement each other: one demonstrates that a practical node can work in the field, while the other shows which platform characteristics make low-power behaviour easier to characterise and optimise.

The observed WAN1310 behaviour should also be attributed carefully. In the analysed outdoor matrix it completes all confirmed-uplink transactions successfully, so the main difference is not reliability failure but higher electrical cost for the same communication task. That difference may arise from modem/stack implementation details, radio timing, and the way the saved run summaries expose TX and RX phases, rather than from any single hardware factor alone. The present results therefore support a claim about measured platform-level behaviour in this setup rather than a blanket claim about intrinsic hardware superiority in all LoRaWAN scenarios.

The study also has clear scope limits: one outdoor site of approximately \SI{100}{\meter}, three repetitions per condition where available, confirmed uplink only, and platform-dependent behaviour under one host-driven workflow. Because the platforms expose some timing events at different instrumentation layers, the paper reports descriptive statistics, run-level validation, and repeated-run variability rather than inferential significance testing. The comparison should therefore be read as a controlled characterisation study, not as a complete generalisation across sites, firmware stacks, or all LoRaWAN modes.

To support reuse, the collected measurement dataset used for the paper results is available on Zenodo at DOI: \href{https://doi.org/10.5281/zenodo.20136328}{10.5281/zenodo.20136328}.

\section{Conclusion}
This paper combines a real agricultural LoRaWAN deployment with controlled platform-level experiments. The deployment evidence shows that multi-year operation is achievable in practice, but only through system-level energy design. The controlled comparison shows that both STM32WL and WAN1310 can execute a complete confirmed-uplink outdoor matrix in this setup, while STM32WL does so with lower energy, lower current, and shorter TX durations across the tested payload and data-rate combinations. Most importantly, the system-level analysis shows that radio energy alone does not explain long lifetime: sleep leakage, sensing strategy, duty cycle, and solar support are the dominant determinants of long-term viability.

{\footnotesize
\bibliographystyle{IEEEtran}
\bibliography{references}
}

\end{document}